\documentclass[twocolumn,showpacs,prd]{revtex4}
\usepackage{graphicx}
\usepackage{color}
\usepackage{amssymb}
\newcommand{\xrm}[1]{{\textstyle \mbox{\rm #1}}}

\newcommand{\abs}[1]{\left| #1\right|}
\begin{document}
\title{First indications of the existence of a 38 MeV light scalar boson.}
\author{Eef van Beveren$^{1}$ and George Rupp$^{2}$}
\affiliation{
$^{1}$Centro de F\'{\i}sica Computacional,
Departamento de F\'{\i}sica, Universidade de Coimbra,
P-3004-516 Coimbra, Portugal\\
$^{2}$Centro de F\'{\i}sica das Interac\c{c}\~{o}es Fundamentais,
Instituto Superior T\'{e}cnico, Technical University of Lisbon,
P-1049-001 Lisboa, Portugal
}
\date{\today}

\begin{abstract}
We present evidence for the existence of a light scalar particle
that most probably couples exclusively to gluons and quarks.
Theoretical and phenomenological arguments are presented
to support the existence of a light scalar boson for confinement
and quark-pair creation.
Previously observed interference effects
allow to set a narrow window for the scalar's mass
and also for its flavor-mass-dependent coupling to quarks.
Here, in order to find a direct signal indicating its production,
we study published BABAR data on leptonic bottomonium decays, viz.\
the reactions
$e^{+}e^{-}$ $\to$ $\pi^{+}\pi^{-}\Upsilon\left( 1,2\,{}^{3\!}S_{1}\right)$
$\to$ $\pi^{+}\pi^{-}e^{+}e^{-}$ (and $\pi^{+}\pi^{-}\mu^{+}\mu^{-}$).
We observe a clear excess signal in the invariant-mass projections
of $e^{+}e^{-}$ and $\mu^{+}\mu^{-}$,
which may be due to the emission of a so far unobserved scalar particle
with a mass of about 38 MeV.
In the process of our analysis, we also find an indication
of the existence of a $b\bar{b}g$ hybrid state at about 10.061 GeV.
Further signals could be interpreted as replicas
with masses two and three times as large as the lightest scalar particle.
\end{abstract}


\maketitle

\section{Introduction}
\label{intro}

In Ref.~\cite{NCA80p401} an $SO(4,2)$ conformally symmetric model
was proposed for strong interactions at low energies,
based on the observation,
published in 1919 by H.~Weyl in Ref.~\cite{AdP364p101},
that the dynamical equations of gauge theories
retain their flat-space-time form
when subject to a conformally-flat metrical field,
instead of the usual Minkowski background.
Confinement of quarks and gluons is then described
through the introduction of two scalar fields which
spontaneously break the $SO(4,2)$ symmetry down to
$SO(3,2)$ and $SO(3)\otimes SO(2)$ symmetry, respectively.
Moreover, a symmetric second-order tensor field is defined
that serves as the metric for flat space-time,
coupling to electromagnetism.
Quarks and gluons, which to lowest order
do not couple to this tensor field,
are confined to an anti-De-Sitter (aDS) universe
\cite{ARXIV07061887},
having a finite radius in the flat space-time.
This way, the model describes quarks and gluons that
oscillate with a universal frequency,
independent of the flavor mass, inside a closed universe,
as well as photons which freely travel through flat space-time.

The fields in the model of Ref.~\cite{NCA80p401}
comprise one real scalar field $\sigma$
and one complex scalar field $\lambda$.
Their dynamical equations were solved in Ref.~\cite{NCA80p401}
for the case that the respective vacuum expectation values,
given by $\sigma_{0}$ and $\lambda_{0}$,
satisfy the relation
\begin{equation}
\abs{\sigma_{0}}\gg \abs{\lambda_{0}}
\;\;\; .
\label{slvacua}
\end{equation}
A solution for $\sigma_{0}$ of particular interest
leads to aDS confinement, via the associated
conformally flat metric given by $\sigma\eta_{\mu\nu}$.

The only quadratic term in the Lagrangian of Ref.~\cite{NCA80p401}
is proportional to
\begin{equation}
-\sigma^{2}\lambda^{\ast}\lambda
\;\;\; .
\label{quadraticterm}
\end{equation}
Hence, under the condition of relation (\ref{slvacua}),
one obtains, after choosing vacuum expectation values,
a light $\sigma$ field, associated with confinement,
and a very heavy complex $\lambda$ field,
associated with electromagnetism.
Weak interactions were not contemplated in Ref.~\cite{NCA80p401},
but one may read electroweak for electromagnetism.
Here, we will study the --- supposedly light --- mass
of the scalar field that gives rise to confinement.

The conformally symmetric model of Ref.~\cite{NCA80p401}
in itself does not easily allow for interactions between hadrons,
as each hadron is described by a closed universe.
Hence, in order to compare the properties of this model
with the actually measured cross sections and branching ratios,
the model has been further simplified,
such that only its main property survives,
namely its flavor-independent oscillations.
This way the full aDS spectrum is, via light-quark-pair creation,
coupled to the channels of two --- or more --- hadronic decay products
for which scattering amplitudes can be measured.

The aDS spectrum reveals itself through the structures
observed in hadronic mass distributions.
However, as we have shown in the past
(see Ref.~\cite{ARXIV10112360} and references therein),
there exists no simple relation between
enhancements in the experimental cross sections
and the aDS spectrum.
It had been studied in parallel, for mesons,
in a coupled-channel model
in which quarks are confined by
a flavor-independent harmonic oscillator
\cite{PRD21p772,PRD27p1527}.
Empirically, based on numerous data on mesonic resonances
measured by a large variety of experimental collaborations,
it was found \cite{ARXIV10091778}
that an aDS oscillation frequency of
\begin{equation}
\omega =190
\;\;\;\xrm{MeV}
\label{uniosc}
\end{equation}
agrees well with the observed results for
meson-meson scattering and meson-pair production
in the light \cite{ZPC30p615},
heavy-light \cite{PRL91p012003},
and heavy \cite{ARXIV10094097} flavor sectors,
thus reinforcing the strategy proposed in Ref.~\cite{NCA80p401}.

Another ingredient of the model for the description
of non-exotic quarkonia, namely the coupling
of quark-antiquark components
to real and virtual two-meson decay channels \cite{AP324p1620}
via $^{3\!}P_{0}$ quark-pair creation,
gives us a clue about the size of the mass of the $\sigma$ field.
For such a coupling it was found that the average radius $r_{0}$
for light-quark-pair creation in quarkonia could be described
by an flavor-independent mass scale, given by
\begin{equation}
M=\frac{1}{2}\omega^{2}\mu r_{0}^{2}
\;\;\; ,
\label{unirad}
\end{equation}
where $\mu$ is the effective reduced quarkonium mass.
In earlier work, the value $\rho_{0}=\sqrt{\mu\omega}r_{0}=0.56$
\cite{PRD21p772,PRD27p1527} was used,
which results in $M=30$ MeV for the corresponding mass scale.
However, the quarkonium spectrum is not very sensitive
to the precise value of the radius $r_{0}$,
in contrast with the resonance widths.
In more recent work \cite{HEPPH0201006,PLB641p265},
slightly larger transition radii have been applied,
corresponding to values around 40 MeV for $M$.
Nevertheless, values of 30--40 MeV
for the flavor-independent mass $M$
do not seem to bear any relation
to an observed quantity for strong interactions.
However, we will next present experimental evidence
for the possible existence of a quantum with a mass of about 38 MeV,
which in the light of its relation to the $^{3\!}P_{0}$ mechanism
we suppose to mediate quark-pair creation.
Moreover, its scalar properties make it a perfect candidate
for the quantum associated with
the above-discussed scalar field for confinement.

\section{Interference}
\label{oscillations}

In Ref.~\cite{PRD79p111501R},
we made notice of an apparent interference effect
around the $D_{s}^{\ast}\bar{D}_{s}^{\ast}$ threshold
in the invariant-mass distribution
of $e^{+}e^{-}\to J/\psi\pi^{+}\pi^{-}$ events,
which we observed in preliminary radiation data
of the BABAR Collaboration \cite{ARXIV08081543}.
The effect, with a periodicity of about 74 MeV,
could be due to interference
between the typical oscillation frequency of 190 MeV
of the $c\bar{c}$ pair and that of the gluon cloud.
\begin{figure}[htbp]
\begin{center}
\begin{tabular}{c}
\includegraphics[width=240pt]{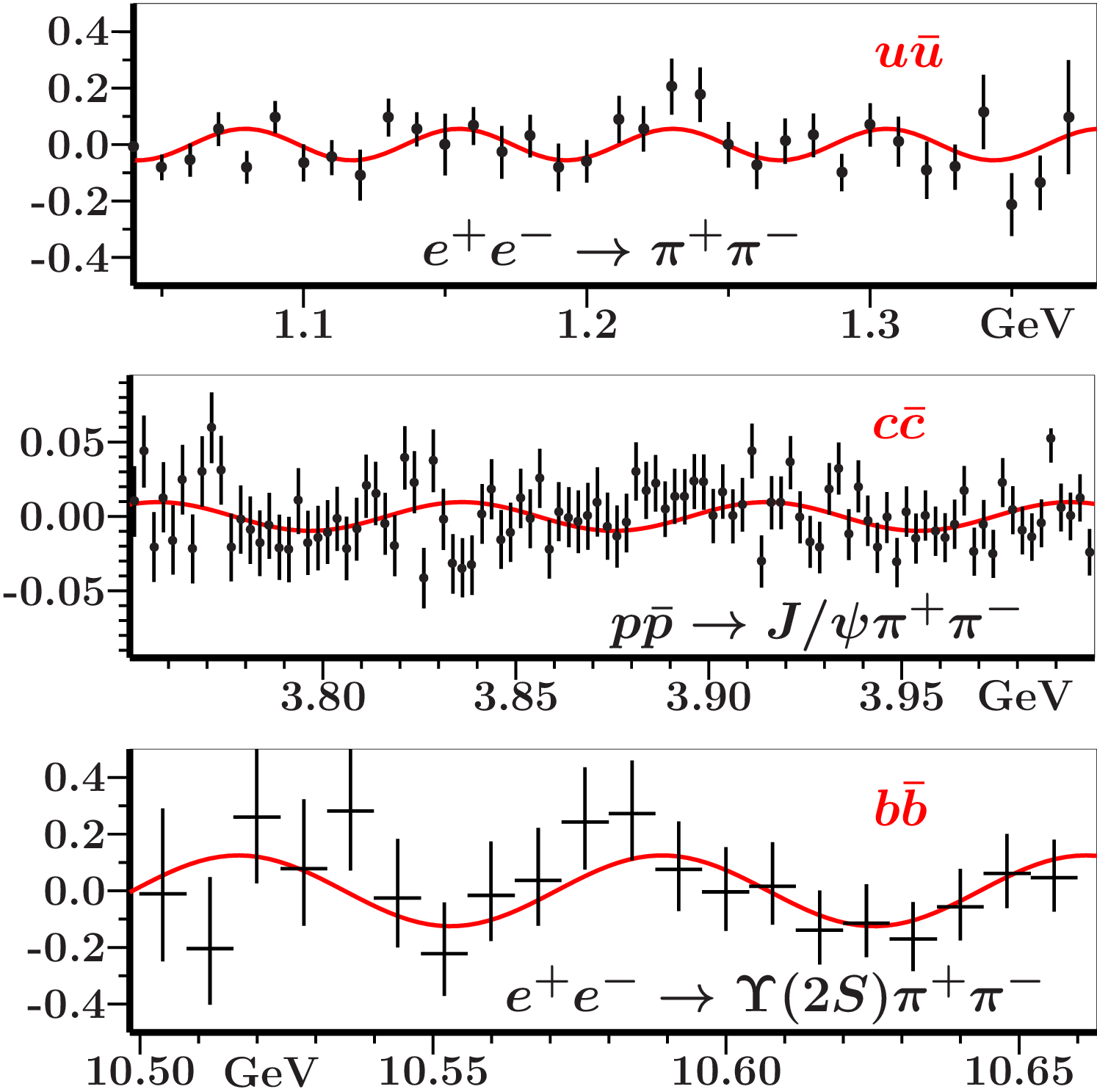}\\ [-10pt]
\end{tabular}
\end{center}
\caption{\small
Fits to the residual data, after subtraction of global fits to:
$e^{+}e^{-}\to\pi^{+}\pi^{-}$ data of the CMD-2 Collaboration
\cite{JETPL82p743}, with a period of 78$\pm$2 MeV
and an amplitude of $\approx$5\% ({\it top});
$p\bar{p}\to J/\psi\pi^{+}\pi^{-}$ data
of the CDF Collaboration \cite{PRL103p152001},
with a period of 79$\pm$5 MeV
and an amplitude of about 0.75\% ({\it middle});
$e^{+}e^{-}\to\Upsilon (2S)\pi^{+}\pi^{-}$ data
of the BABAR Collaboration \cite{PRD78p112002},
with a period of 73$\pm$3 MeV
and an amplitude of some 12.5\% ({\it bottom}).
}
\label{interference}
\end{figure}
Later, in Ref.~\cite{ARXIV10095191},
we reported evidence for small oscillations
in electron-positron and proton-antiproton annihilation data,
with a periodicity of 76$\pm$2 MeV, independent of the beam energy.
The latter observations are summarized in Fig.~\ref{interference}.

Amongst the various scenarios to explain the phenomenon
presented in Ref.~\cite{ARXIV10095191},
one was rather intriguing, namely the postulated existence of
gluonic oscillations, possibly surface oscillations,
with a frequency of about 38 MeV. These would then, upon interfering
with the universal quarkonia frequency $\omega =190$ MeV
\cite{PRD21p772,PRD27p1527},
lead to the observed oscillations.

In the present work,
we are going to further elaborate on the hypothesis
that the observed oscillations are caused by
slow gluonic oscillations.
However, we will find instead that the phenomenon
are more likely to be associated with the
interquark exchange of a scalar particle
with a mass of about 38 MeV.
Moreover, from the fact that the observed oscillations
are more intense for bottomonium than for light quarks,
we assume that the coupling of this light scalar to quarks
increases with the quark mass.
It seems to correspond well to the scalar particle
of the model of Ref.~\cite{NCA80p401},
and to the enigmatic mass parameter $M$ related to
the $^{3\!}P_{0}$ pair-creation mechanism.

\section{Signs of light scalar particle?}
\label{phenomenon}

In Ref.~\cite{PRD78p112002}, the BABAR Collaboration
presented an analysis of data on
$e^{+}e^{-}$ $\to$ $\pi^{+}\pi^{-}
\Upsilon\left( 1,2\,{}^{3\!}S_{1}\right)$
$\to$ $\pi^{+}\pi^{-}\ell^{+}\ell^{-}$
($\ell =e$ and $\ell =\mu$),
with the aim to study hadronic transitions between
$b\bar{b}$ excitations
and the $\Upsilon\left( 1\,{}^{3\!}S_{1}\right)$
and $\Upsilon\left( 2\,{}^{3\!}S_{1}\right)$,
based on 347.5 fb$^{-1}$ of data
taken with the BABAR detector at the PEP-II storage rings.

In Ref.~\cite{ARXIV10094097}, we reported evidence
for the existence of the
$\Upsilon\left( 2\,{}^{3\!}D_{1}\right)$
at about 10.495 GeV,
and some indications of the existence of the
$\Upsilon\left( 1\,{}^{3\!}D_{1}\right)$
at about 10.098 GeV,
by analyzing the above BABAR data.
In the present paper, these data
are further analyzed.

\begin{figure}[htpb]
\begin{center}
\begin{tabular}{c}
\includegraphics[width=240pt]{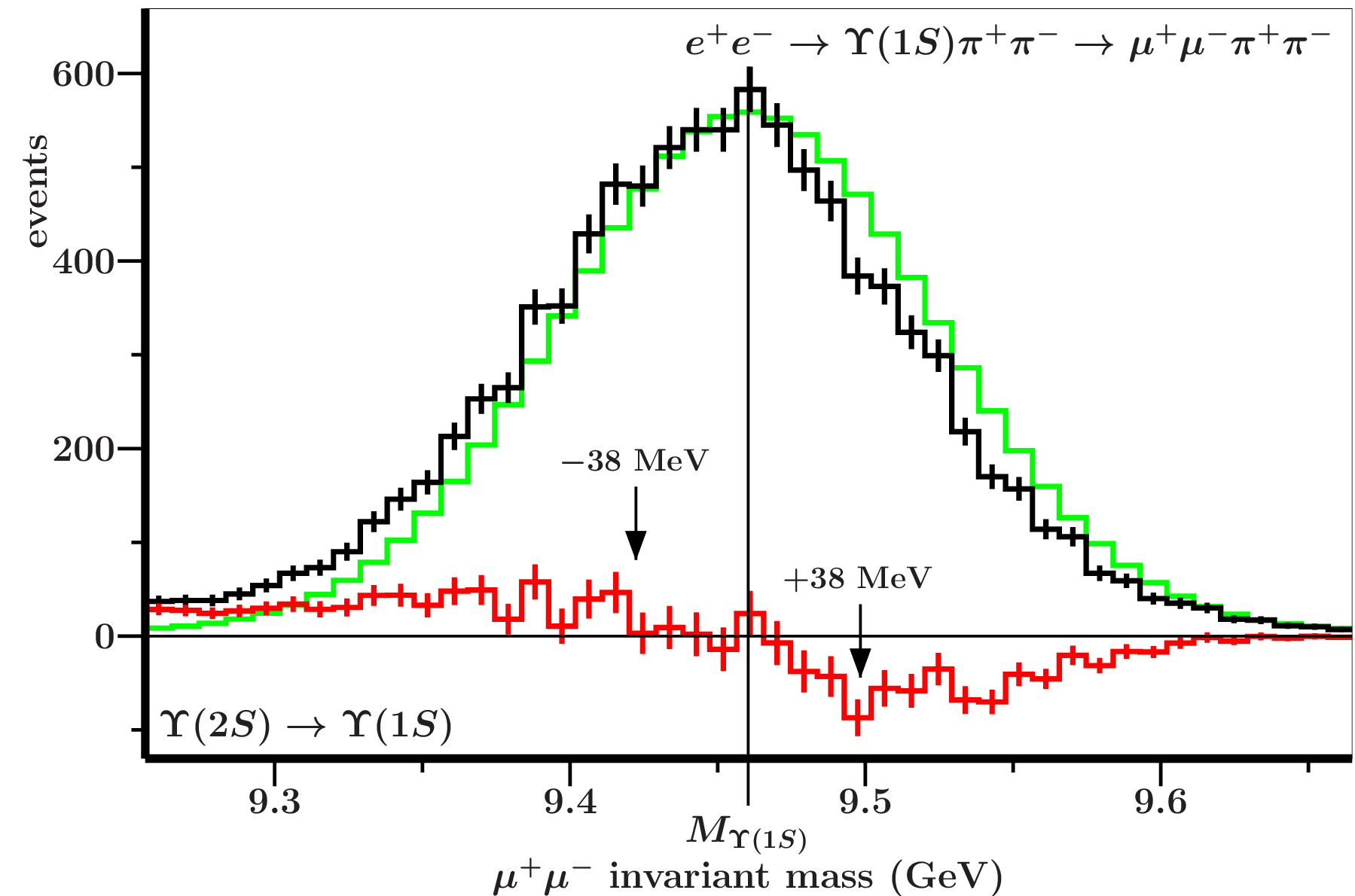}\\ [-10pt]
\end{tabular}
\end{center}
\caption{\small
Invariant $\mu^{+}\mu^{-}$ mass distribution for events
identified as stemming from the reaction
$e^{+}e^{-}$ $\to$ $\Upsilon\left( 2\,{}^{3\!}S_{1}\right)$ $\to$
$\pi^{+}\pi^{-}\Upsilon\left( 1\,{}^{3\!}S_{1}\right)$
$\to$ $\pi^{+}\pi^{-}\mu^{+}\mu^{-}$.
Data (black) are taken from Ref.~\cite{PRD78p112002}.
The bin size equals 9 MeV.
Statistical errors are shown by vertical bars.
The vertical line indicates
$M_{\mu^{+}\mu^{-}}=M_{\Upsilon\left( 1\,{}^{3\!}S_{1}\right)}$.
The Gaussian distribution (gray, green in online version)
and the excess data at the bottom of the figure
(black, red in online version) are explained in the text.
}
\label{mumuAll2S}
\end{figure}

The selection procedure for the data is well described by BABAR
in Refs.~\cite{PRD78p112002,PRL104p191801,ARXIV09100423}.
In Fig.~\ref{mumuAll2S}, we study
the invariant-mass distribution of muon pairs
obtained from the BABAR data set \cite{PRD78p112002}
for the reaction $e^{+}e^{-}$ $\to$
$\Upsilon\left( 2\,{}^{3\!}S_{1}\right)$ $\to$
$\pi^{+}\pi^{-}\Upsilon\left( 1\,{}^{3\!}S_{1}\right)$
$\to$ $\pi^{+}\pi^{-}\mu^{+}\mu^{-}$,
and for a bin size equal to 9 MeV.

Furthermore, we show in Fig.~\ref{mumuAll2S}
a simple Gaussian distribution
(gray histogram, green in online version),
with a width of 89 MeV, around the
$\Upsilon\left( 1\,{}^{3\!}S_{1}\right)$ peak.
We observe that, with respect to the Gaussian distribution,
there is an excess of data
for $M_{\mu^{+}\mu^{-}}$ below
the $\Upsilon\left( 1\,{}^{3\!}S_{1}\right)$ mass,
and a deficit of data
for $M_{\mu^{+}\mu^{-}}$ thereabove.
Actually, we have chosen the Gaussian distribution such
that the total difference between
the data under the Gaussian histogram
and the experimental data vanishes.
The excess signal is indicated
(dark, shaded, histogram, red in online version)
at the bottom of Fig.~\ref{mumuAll2S}.

We observe from Fig.~\ref{mumuAll2S}
that the excess of data sets out for masses
some 40 MeV below
the $\Upsilon\left( 1\,{}^{3\!}S_{1}\right)$ mass,
viz.\ at about $M_{\mu^{+}\mu^{-}}=9.42$ GeV, and then
towards lower $\mu^{+}\mu^{-}$ injvariant masses,
leaving a small signal on top of the increasing background tail,
up to about 9.33 GeV.
The deficit data exhibit enhancements at about
$M_{\mu^{+}\mu^{-}}=9.50$, 9.54 and 9.57 GeV,
i.e., 38, 76, and 114 MeV above
the $\Upsilon\left( 1\,{}^{3\!}S_{1}\right)$ mass, respectively.

\begin{figure}[htpb]
\begin{center}
\begin{tabular}{c}
\raisebox{10pt}{\bf a}\includegraphics[width=230pt]{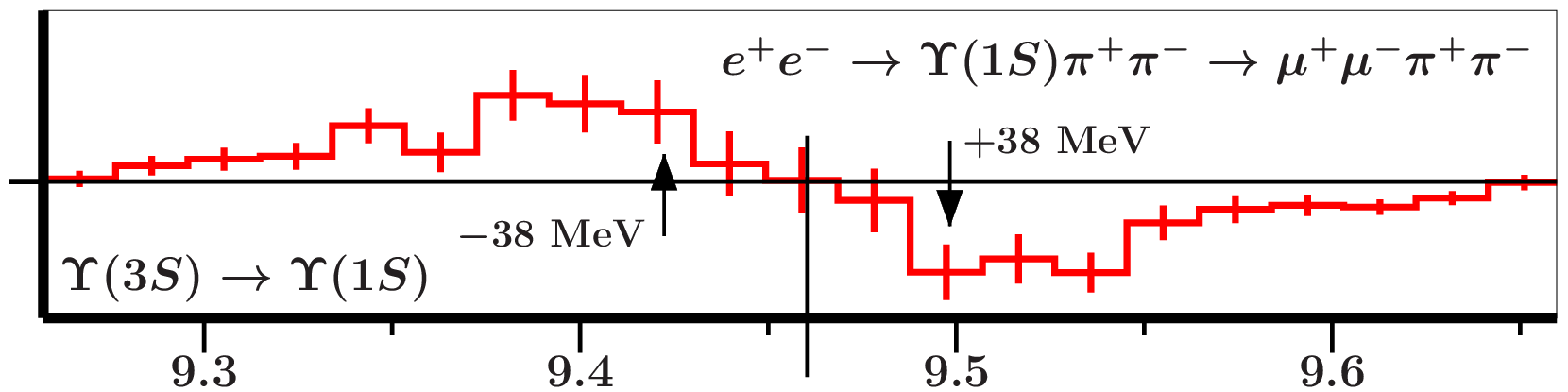}\\
\raisebox{10pt}{\bf b}\includegraphics[width=230pt]{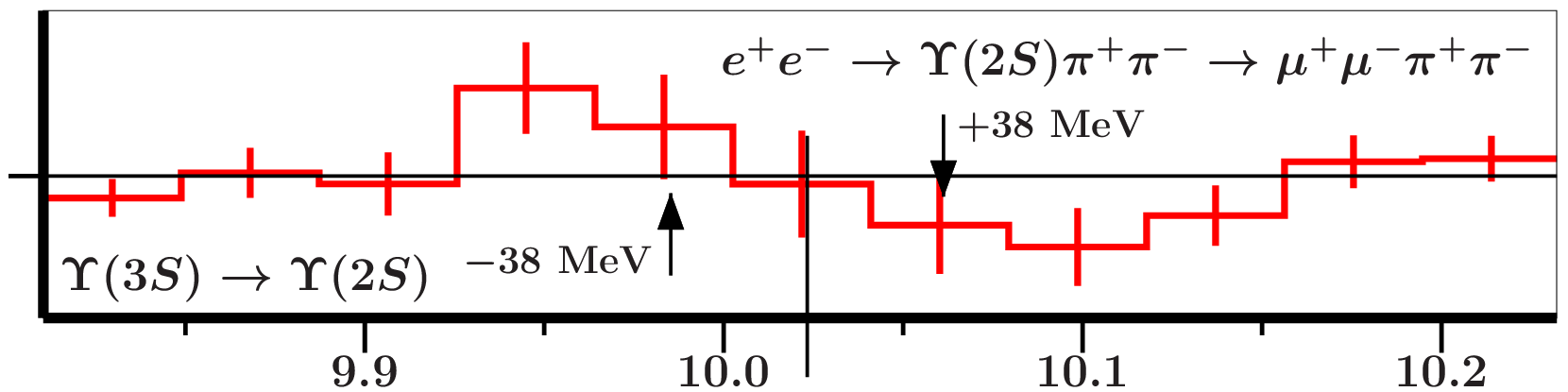}\\
\raisebox{10pt}{\bf c}\includegraphics[width=230pt]{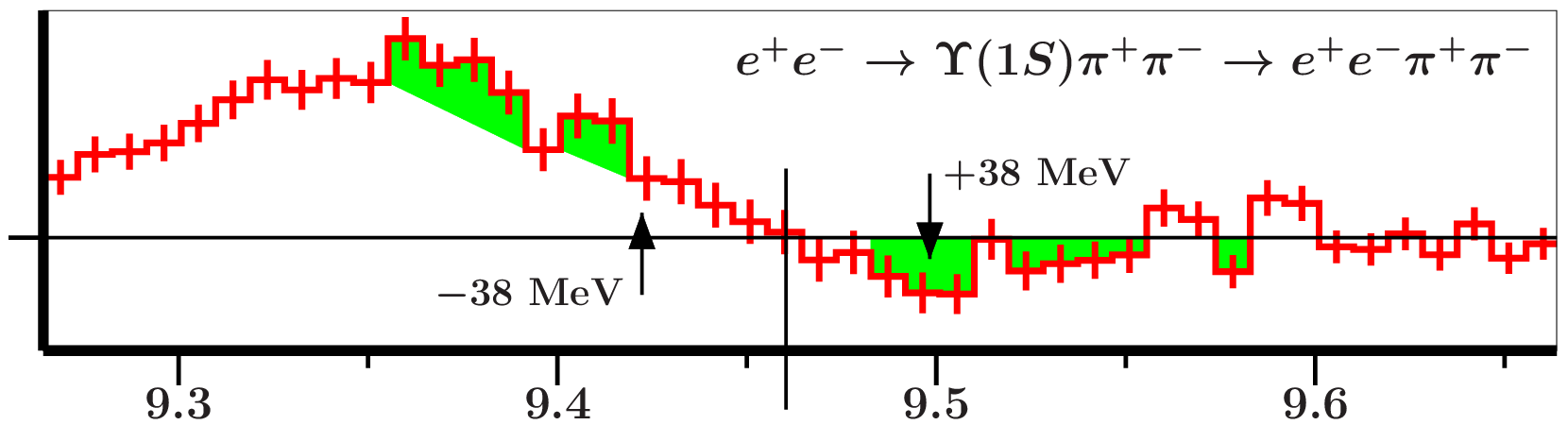}\\ [-10pt]
\end{tabular}
\end{center}
\caption{\small
Excess signals
in invariant $\mu^{+}\mu^{-}$ (a, b)
and $e^{+}e^{-}$ (c) mass distributions:
for the reactions
$\Upsilon\left( 3\,{}^{3\!}S_{1}\right)$ $\to$
$\pi^{+}\pi^{-}\Upsilon\left( 1\,{}^{3\!}S_{1}\right)$
$\to$ $\pi^{+}\pi^{-}\mu^{+}\mu^{-}$,
using bins of 19 MeV (a);
$\Upsilon\left( 3\,{}^{3\!}S_{1}\right)$ $\to$
$\pi^{+}\pi^{-}\Upsilon\left( 2\,{}^{3\!}S_{1}\right)$
$\to$ $\pi^{+}\pi^{-}\mu^{+}\mu^{-}$,
using bins of 38 MeV (b);
$e^{+}e^{-}$ $\to$
$\pi^{+}\pi^{-}\Upsilon\left( 1\,{}^{3\!}S_{1}\right)$
$\to$ $\pi^{+}\pi^{-}e^{+}e^{-}$
for all available data,
using bins of 9 MeV (c).
Statistical errors are shown by vertical bars.
The vertical lines indicate
$M_{\mu^{+}\mu^{-}}=M_{\Upsilon\left( 1,2\,{}^{3\!}S_{1}\right)}$.
Horizontal units,
for $M_{\mu^{+}\mu^{-}}$ (a, b)
and $M_{e^{+}e^{-}}$ (c),
are given in GeV.
}
\label{alldiff}
\end{figure}

In Fig.~\ref{alldiff},
we have collected excess signals for other reactions,
thereby following similar procedures as before.
We have selected all reactions
with some reasonable statistics
from BABAR \cite{PRD78p112002} data,
viz.\
$\Upsilon\left( 3\,{}^{3\!}S_{1}\right)$ $\to$
$\pi^{+}\pi^{-}\Upsilon\left( 1\,{}^{3\!}S_{1}\right)$
$\to$ $\pi^{+}\pi^{-}\mu^{+}\mu^{-}$
(Fig.~\ref{alldiff}a),
$\Upsilon\left( 3\,{}^{3\!}S_{1}\right)$ $\to$
$\pi^{+}\pi^{-}\Upsilon\left( 2\,{}^{3\!}S_{1}\right)$
$\to$ $\pi^{+}\pi^{-}\mu^{+}\mu^{-}$
(Fig.~\ref{alldiff}b),
and
$e^{+}e^{-}$ $\to$
$\pi^{+}\pi^{-}\Upsilon\left( 1\,{}^{3\!}S_{1}\right)$
$\to$ $\pi^{+}\pi^{-}e^{+}e^{-}$
for all available data
(Fig.~\ref{alldiff}c).
The data binning has been chosen
in order to optimize statistics.

In Fig.~\ref{alldiff}a, which is 19 MeV binned,
we observe two connected enhancements at 38 and 76 MeV
below the $\Upsilon\left( 1\,{}^{3\!}S_{1}\right)$ mass,
and a third one, 38 MeV further downwards.
Above the $\Upsilon\left( 1\,{}^{3\!}S_{1}\right)$ mass,
we observe two connected negative enhancements,
38 and 76 MeV higher up in mass.
In Fig.~\ref{alldiff}b, which is 38 MeV binned,
we observe two connected enhancements at 38 and 76 MeV
below the $\Upsilon\left( 2\,{}^{3\!}S_{1}\right)$ mass,
and three connected enhancements at 38, 76 and 114 MeV
above the $\Upsilon\left( 2\,{}^{3\!}S_{1}\right)$ mass.

In itself,
it is not surprising that an intrinsic asymmetry
in mass distributions leads to excess on one side of the center
and to a deficit on the other side,
with respect to a symmetric distribution.
However, we do observe structure in the excess and deficit data.
This can be most clearly seen in the excess distribution
of the reaction $e^{+}e^{-}$ $\to$
$\pi^{+}\pi^{-}\Upsilon\left( 1\,{}^{3\!}S_{1}\right)$
$\to$ $\pi^{+}\pi^{-}e^{+}e^{-}$ (see Fig.~\ref{alldiff}c),
where we do not opt for an overall vanishing excess,
as we did for the other reactions.
The excess signal below
$M_{e^{+}e^{-}}=M_{\Upsilon\left( 1\,{}^{3\!}S_{1}\right)}$
is mainly due to Bremsstrahlung,
as explained by BABAR in Ref.~\cite{PRD78p112002}.
Nevertheless, on top of the Bremsstrahlung background
one observes something extra in the invariant-mass interval
9.35--9.42 GeV.
No doubt, it is an additional signal of hardly more than 1$\sigma$.
However, it is in roughly the same invariant-mass interval
where we find some excess signal in the other three reactions.
Moreover, the pronounced deficit at about 9.50 GeV
comes out also in the same invariant-mass interval
where we find a deficit signal in the other three reactions.
The deficit enhancements at around 9.54 and 9.575 GeV
are hardly distinguishable from zero,
but show up in the expected energy intervals
of 76 and 114 MeV above
the $\Upsilon\left( 1\,{}^{3\!}S_{1}\right)$ mass.

In Ref.~\cite{PRL103p081803},
the BABAR Collaboration studied evidence for a light scalar boson in
$99\times 10^{6}$ $\Upsilon\left( 2\,{}^{3\!}S_{1}\right)$
and $122\times 10^{6}$ $\Upsilon\left( 3\,{}^{3\!}S_{1}\right)$
radiative decays,
$\Upsilon\left( 2,3\,{}^{3\!}S_{1}\right)$
$\to$ $\gamma A^{0}$ $\to$ $\gamma\mu^{+}\mu^{-}$,
as suggested by extensions of the standard model,
in which a light $CP$-odd Higgs boson $A^{0}$
naturally couples strongly to $b$ quarks
\cite{PLB120p346,PLB222p11,PLB237p307,PRD39p844,PRL98p081802}.
BABAR reported \cite{PRL103p081803}
to have found no evidence for such processes in the mass range
$0.212\le m_{A^{0}}\le 9.3$ GeV.
Furthermore,
in Ref.~\cite{PRL104p151802}
charged-lepton flavor-violating processes
were studied by BABAR for similar large amounts of data,
while BABAR tested lepton universality
in Ref.~\cite{PRL104p191801} also for about $10^{8}$
$\Upsilon\left( 1\,{}^{3\!}S_{1}\right)$
$\to$ $\ell^{+}\ell^{-}$ decay events.
No significant deviations from the Standard-Model
expectations were observed in these experiments.

\begin{figure}[htpb]
\begin{center}
\begin{tabular}{c}
\includegraphics[width=240pt]{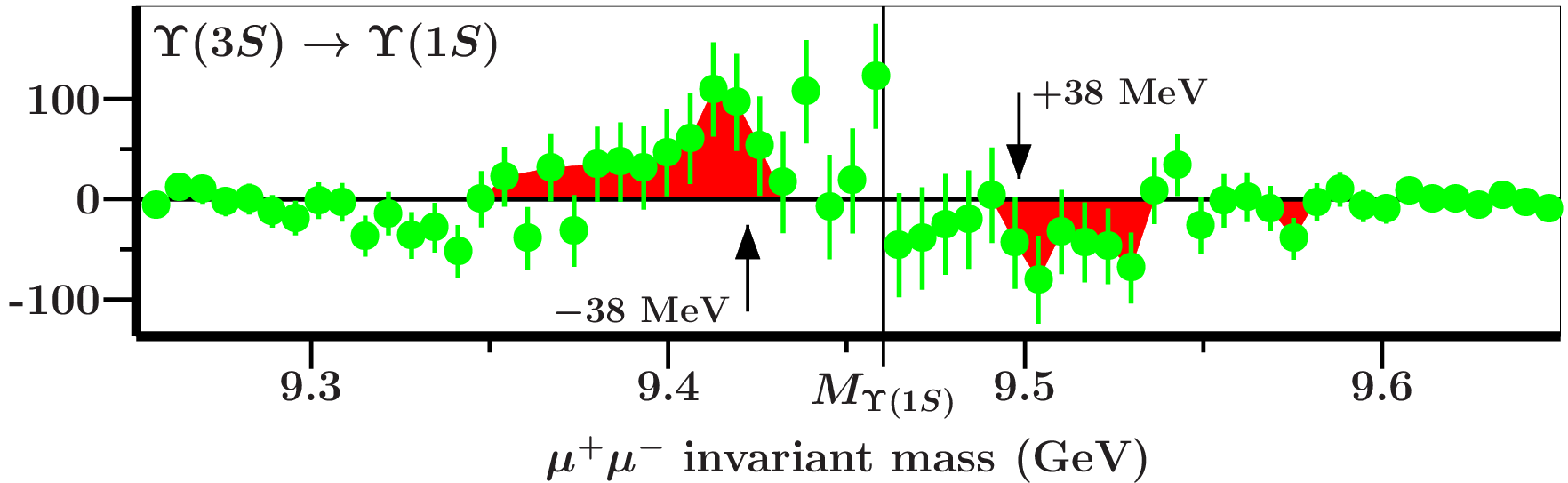}\\ [-10pt]
\end{tabular}
\end{center}
\caption{\small
Event distribution of the excess signal
taken from Ref.~\cite{ARXIV09100423},
in the invariant-$\mu^{+}\mu^{-}$-mass distribution
for the reaction
$\Upsilon\left( 2\,{}^{3\!}S_{1}\right)$ $\to$
$\pi^{+}\pi^{-}\Upsilon\left( 1\,{}^{3\!}S_{1}\right)$
$\to$ $\pi^{+}\pi^{-}\mu^{+}\mu^{-}$,
using bins of 6.5 MeV.
Statistical errors are shown by vertical bars.
The shaded areas (dark, red in online version)
are discussed in the text.
The vertical line indicates
$M_{\mu^{+}\mu^{-}}=M_{\Upsilon\left( 1\,{}^{3\!}S_{1}\right)}$.
}
\label{elisa}
\end{figure}

With such amounts of events, the BABAR Collaboration should
be able to confirm or invalidate our observations.
Therefore, it was a pleasure for us to find
a similar graph for the residual data
in their presentation
by E.~Guido on behalf of the BABAR Collaboration
\cite{ARXIV09100423}.
This result is shown in Fig.~\ref{elisa}.
Moreover, the analysis in Ref.~\cite{ARXIV09100423}
took all known possible origins of asymmetry into account.
Consequently, what is left (see Fig.~\ref{elisa})
cannot be explained by known physics.

When we study Fig.~\ref{elisa} in more detail,
we observe a clear structure just around and also below
9.42 GeV, exactly as we found above.
Furthermore, we see three minima in the data deficit
above
$M_{\mu^{+}\mu^{-}}=M_{\Upsilon\left( 1\,{}^{3\!}S_{1}\right)}$,
namely at about 9.50, 9.53, and 9.57 GeV,
i.e., 40, 70, and 115 MeV above
the
$\Upsilon\left( 1\,{}^{3\!}S_{1}\right)$ mass,
respectively.
Again, these minima do not have more significance than 2$\sigma$,
but they are compatible with the values obtained before.

Moreover, Ref.~\cite{ARXIV09100423} confirms
our assumption that for $\mu^{+}\mu^{-}$ background is small.
Also, it states that the, here reported,
systematic uncertainties due to the differences
between data and simulation in the processes
$\Upsilon\left( 1\,{}^{3\!}S_{1}\right)$ $\to$ $\tau^{+}\tau^{-}$
and
$\Upsilon\left( 1\,{}^{3\!}S_{1}\right)$ $\to$ $\mu^{+}\mu^{-}$
cancel, at least in part, in their ratio. This
implies that a similar excess is found
in the
$\Upsilon\left( 1\,{}^{3\!}S_{1}\right)$ $\to$ $\tau^{+}\tau^{-}$
decay.

\section{Hypothesis of a light scalar particle}

When the $\Upsilon\left( 1\,{}^{3\!}S_{1}\right)$
decays into a lepton pair and an additional light scalar particle,
the invariant mass of the lepton pair will be smaller
than the $\Upsilon\left( 1\,{}^{3\!}S_{1}\right)$ mass.
Depending on the amount of momentum taken
by the light scalar particle,
the event will pass through BABAR's selection procedure.

Let us designate the hypothetical scalar boson by $E(38)$.
If we assume its mass to be about 38 MeV,
the mentioned lepton pair will have invariant masses that are
smaller the $\Upsilon\left( 1\,{}^{3\!}S_{1}\right)$ mass by at
least 38 MeV. Hence, we will find an excess of events
for invariant $\ell^{+}\ell^{-}$ masses below about 9.42 GeV,
with a spreading comparable to
the spreading of the data without $E(38)$ production.
This seems to agree with what we observed
in Sec.~\ref{phenomenon}.
Furthermore, when several $E(38)$'s are produced,
such signals will repeat themselves, viz.\
at 76, 114, \ldots\ MeV
below the $\Upsilon\left( 1\,{}^{3\!}S_{1}\right)$ mass,
which also seems to agree with what we found
in Sec.~\ref{phenomenon}.

\begin{figure}[htpb]
\begin{center}
\begin{tabular}{c}
\includegraphics[width=220pt]{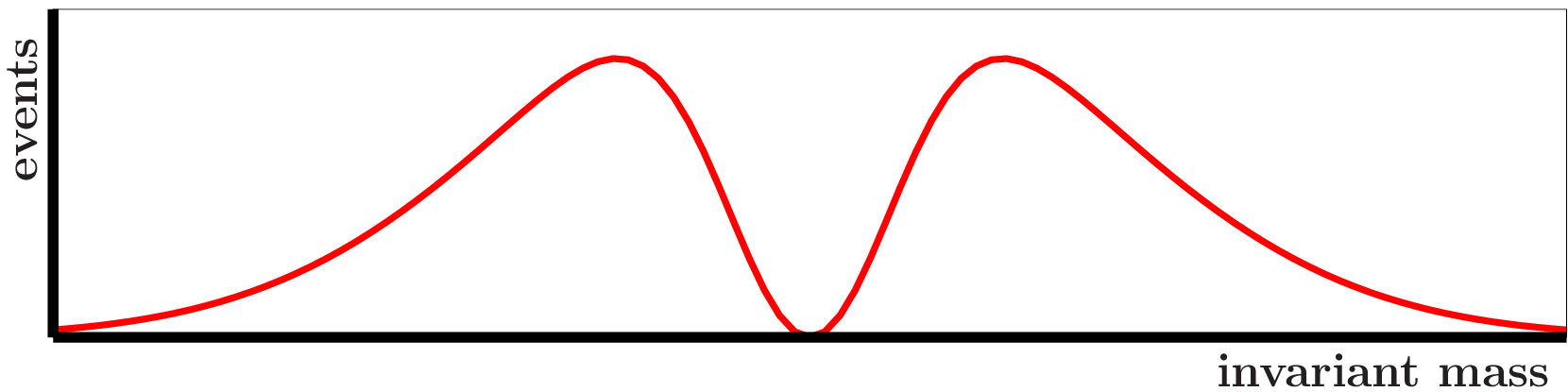}\\ [-10pt]
\end{tabular}
\end{center}
\caption{\small
Signal shape stemming from the hybrid transitions.
}
\label{dip}
\end{figure}

In order to explain the structures in the deficit signal,
we must assume that the $E(38)$ can be loosely bound
inside a $b\bar{b}$ state, giving rise to a kind of hybrid.
For such a  situation, we have two possibilities, viz.\
either hybrid-to-hybrid transitions
$\Upsilon '\left( 3,2\,{}^{3\!}S_{1}\right)$ $\to$
$\pi^{+}\pi^{-}\Upsilon '\left( 2,1\,{}^{3\!}S_{1}\right)$
(the primes indicate hybrids),
or hybrid-to-$b\bar{b}$ transitions
$\Upsilon '\left( 3,2\,{}^{3\!}S_{1}\right)$ $\to$
$\pi^{+}\pi^{-}\Upsilon\left( 2,1\,{}^{3\!}S_{1}\right)$.
The events stemming from the former transitions
come in the initial-state band,
since the mass difference of the two hybrids
equals the mass difference of the pure $b\bar{b}$ states.
However, the hybrid-to-$b\bar{b}$ events do not end up
in the initial-state band of events,
since the mass difference is about 38 MeV too large.
Hence, the deficit can be due to those events
that were expected to come with a
higher invariant two-lepton mass,
but ended up outside the initial-state band
(see Fig.~\ref{dip}).
The dips at about 9.50 and 9.57 GeV
in Fig.~\ref{elisa}
certainly confirm such shapes, whereas
the one in the middle is more distorted.
Furthermore, the shapes of the three dips
suggest that, actually, the higher mass phenomena
are rather related to replicas \cite{PRD65p085026} of the $E(38)$
with masses that are two and three times heavier than the $E(38)$,
than to signals stemming from
the production of two and three $E(38)$'s.

\begin{figure}[htpb]
\begin{center}
\begin{tabular}{c}
\includegraphics[width=230pt]{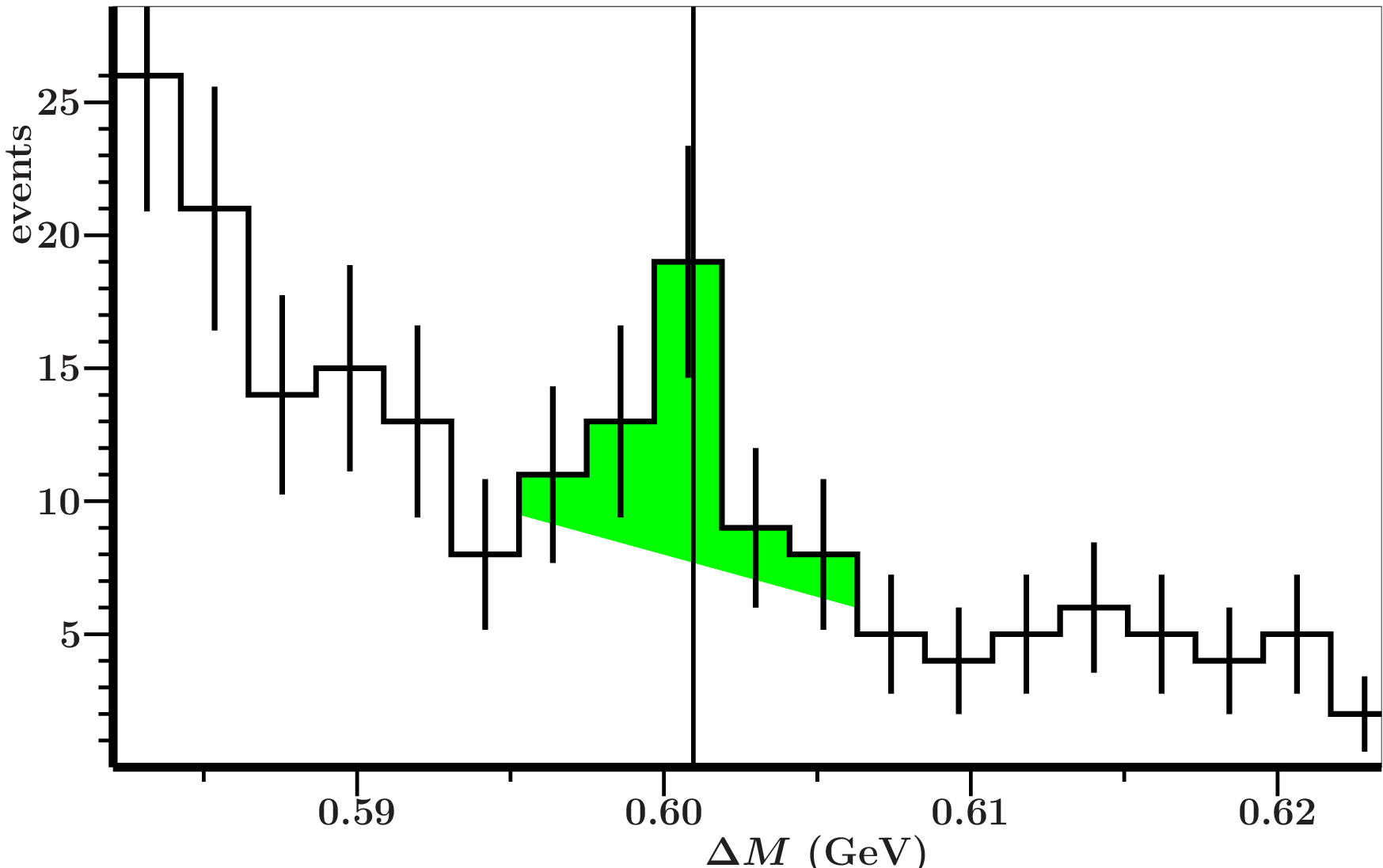}\\ [-10pt]
\end{tabular}
\end{center}
\caption{\small
Possible sign of the $\Upsilon '\left( 2\,{}^{3\!}S_{1}\right)$ hybrid.
The vertical line indicates where
$M_{\Upsilon\left( 2\,{}^{3\!}S_{1}\right)}+38$ MeV comes out,
in terms of $\Delta M=M_{\pi^{+}\pi^{-}\mu^{+}\mu^{-}}-M_{\mu^{+}\mu^{-}}$.
Data are from Ref.~\cite{PRD78p112002}.
}
\label{hybrid}
\end{figure}

Hence, we have to search for possible hybrids
\cite{PREP40p75,ZPC26p307,PREP454p1,HEPPH0211289,HEPPH0309228}.
In Fig.~\ref{hybrid}, we show the event distribution
for the invariant mass $\Delta M$, which is defined
\cite{PRD78p112002} by $\Delta M=$
$M_{\pi^{+}\pi^{-}\mu^{+}\mu^{-}}-M_{\mu^{+}\mu^{-}}$,
where the latter mass is supposed
to be the $\Upsilon\left( 1\,{}^{3\!}S_{1}\right)$ mass.
Thus, a signal with the shape of a narrow Breit-Wigner resonance
seems to be visible on the slope of the
$\Upsilon\left( 2\,{}^{3\!}S_{1}\right)$
resonance, though with little more than 2$\sigma$ relevance.
Nevertheless, by coincidence or not, it comes out exactly in the
expected place, namely at
$M_{\Upsilon\left( 2\,{}^{3\!}S_{1}\right)}+38$ MeV.
Unfortunately, the data \cite{PRD78p112002}
do not have enough statistics to pinpoint
possible higher excitations as well.
So we cannot relate, to a minimum degree of accuracy,
the other observed deficit enhancements
to the possible existence of hybrid states.

\section{Scalar Glueball}
\label{gluons}

From the fact that the $E(38)$ has not been observed before,
one must conclude that it probably does not interact --- at least
to leading order --- through electroweak forces,
but instead couples exclusively to quarks and gluons.
The interference effect
we discussed in Sec.~\ref{oscillations}
might well be explained by $E(38)$ exchange between the quarks,
which interferes with the natural quark oscillations.
Moreover, since the interference effect is smaller for light quarks
than for heavier ones, it is likely that their coupling to the $E(38)$
is proportional to flavor mass, as one expects from theory.

The $E(38)$ could very well be
just a light scalar glueball,
albeit much lighter than found in Refs.
\cite{HEPPH0209225,PRL79p1998,HEPPH9905250,PRD60p034509,HEPLAT0510066,
PREP454p1,PRD74p034019,ARXIV07111435}.
Its low mass precludes decay into hadrons,
while the absence of electroweak couplings
does not allow it to decay into leptons either, at
least to lowest order.
It may decay, though, into photons via virtual quark loops,
and through photons, eventually, into $e^{+}e^{-}$ pairs.
However, the probability for such reactions to occur
is extremely remote,
since the coupling between the quarks and the light scalar
is proportional to the quark mass \cite{NCA80p401}.
In the case of bottom quarks, we found events of the order of or less
than one percent of the total.
For light quarks, their mass ratio with respect to bottom quarks
reduces this rate to $10^{-7}$--$10^{-6}$.

Hadrons will certainly interact with a light scalar ball of glue.
For example, a proton struck by such a scalar particle
may absorp it and then emit photons,
or decay into a neutron and a lepton-neutrino pair.
Yet another possibility is that, being closed universes themselves,
these scalar particles mainly collide elastically with hadronic matter.
In that case, depending on their linear momentum,
they may remove light nuclei from atoms.
Nevertheless, we do not see such processes happening around us.
Therefore, these light scalars
are probably not abundantly present near us.
However, in the early universe they may have existed,
most probably inflated to hadrons under collisions.
On the other hand, interactions with hadrons might have been observed
in bubble-chamber experiments, where isolated protons could be the result
of collisions with a light scalar particle
emerging from one of the interaction vertices.

Light Higgs fields have been considered in supersymmetric extensions of
the Standard Model \cite{PRL95p041801}. Furthermore, axions,
which appear in models motivated by astrophysical observations,
are assumed to have Higgs-like couplings \cite{PRD79p075008}.
Model predictions for the branching fraction
of $\Upsilon\to\gamma+$Higgs decays,
for Higgs masses below $2m_{b}$ \cite{PRD73p111701R},
range from $10^{-6}$ \cite{PRD79p075008,PRD76p051105R}
to $10^{-4}$ \cite{PRD76p051105R}.
Furthermore, the three anomalous events
observed in the HyperCP experiment \cite{PRL94p021801}
were interpreted as the production of a scalar boson
with a mass of 214.3 MeV, decaying into a pair of muons
\cite{PRL98p081802,MPLA22p1373}.
However, in Ref.~\cite{PRL103p081803}
the BABAR Collaboration found no evidence for
dimuon decays of a light scalar particle
in radiative decays of
$\Upsilon\left( 2\,{}^{3\!}S_{1}\right)$
and
$\Upsilon\left( 3\,{}^{3\!}S_{1}\right)$
mesons.
The BABAR limits for
dimuon decays of a light scalar particle
rule out much of the parameter space allowed
by the light-Higgs \cite{PRD76p051105R}
and axion \cite{PRD79p075008} models.
Nonetheless, in Ref.~\cite{ARXIV09111766}
Y.-J.~Zhang and H.-S.~Shao,
pointed out that the transitions
$\Upsilon\to\ell^{+}\ell^{-}+$Higgs
are not yet excluded by the
lepton-universality test in
$\Upsilon\left( 1\,{}^{3\!}S_{1}\right)$ decays
studied by BABAR in
Refs.~\cite{ARXIV09100423,PRL104p191801}.

The light scalar glueball we have discussed here
seems to correspond to
the lowest-order empty-universe solution
of Ref.~\cite{NCA80p401} for strong interactions.
It has similar properties as the electroweak Higgs,
but now for strong interactions.
Quarks couple to it with an intensity which is proportional
to their mass, in the same way
that mass couples to gravity \cite{SPAWB1915p844}.
In the Standard Model
\cite{NP22p579,PRL19p1264,StandardModel,PRD2p1285},
the Higgs boson and the graviton
are the only particles yet to be observed,
and no Higgs particle for strong interactions is anticipated.
However, N.~T\"{o}rnqvist recently proposed
\cite{HEPPH0204215,PLB619p145} the light scalar-meson nonet \cite{ZPC30p615}
as the Higgs bosons of strong interactions,
while in Ref.~\cite{PLB619p145} he obtained a nonzero pion mass
by means of a small breaking of a relative symmetry
between the electroweak and the strong interactions.
A relation between the lightest scalar-meson nonet
and glueballs has often been advocated
by P.~Minkowski and W.~Ochs (see e.g.\ Ref.~\cite{EPJC9p283}).
In our view though, the light scalar mesons are dynamically generated
through $q\bar{q}$ pair creation/annihilation,
which mixes the quark-antiquark and dimeson sectors \cite{ZPC30p615}.

\section{Epilogue}

In Sec.~\ref{intro}
we discussed why we expect an additional scalar particle to exist,
besides the Higgs boson for the electroweak sector.
Furthermore, we have estimated its mass based on the average radius
for $^{3\!}P_{0}$ quark-pair creation,
which had been extracted over the past three decades
from numerous data on mesonic resonances
(see Ref.~\cite{ARXIV10112360} and references therein).
In Sec.~\ref{oscillations}
we recalled our results on an apparent interference effect
in annihilation data,
and stressed the possibility
that it may stem from some internal oscillation
with a frequency of about 38 MeV.
In Sec.~\ref{phenomenon}
we showed that missing data in the reactions
$e^{+}e^{-}$ $\to$ $\pi^{+}\pi^{-}\Upsilon\left( 1\,{}^{3\!}S_{1}\right)$
$\to$ $\pi^{+}\pi^{-}e^{+}e^{-}$
and
$e^{+}e^{-}$ $\to$ $\pi^{+}\pi^{-}\Upsilon\left( 1\,{}^{3\!}S_{1}\right)$
$\to$ $\pi^{+}\pi^{-}\mu^{+}\mu^{-}$
exhibit maxima at $M$, $2M$, and $3M$,
for $M\approx 38$ MeV.

Each of the results, viz.\
the interference effect observed in
Ref.~\cite{PRD79p111501R},
the small flavor-independent oscillations
in electron-positron and proton-antiproton annihilation data,
observed in Ref.~\cite{ARXIV10095191}
and summarized in Fig.~\ref{interference},
the excess signals visible in
the $\mu^{+}\mu^{-}$ mass distributions of
$\Upsilon\left( 2\,{}^{3\!}S_{1}\right)$ $\to$
$\pi^{+}\pi^{-}\Upsilon\left( 1\,{}^{3\!}S_{1}\right)$
$\to$ $\pi^{+}\pi^{-}\mu^{+}\mu^{-}$
(Fig.~\ref{mumuAll2S}),
in $\Upsilon\left( 3\,{}^{3\!}S_{1}\right)$ $\to$
$\pi^{+}\pi^{-}\Upsilon\left( 1\,{}^{3\!}S_{1}\right)$
$\to$ $\pi^{+}\pi^{-}\mu^{+}\mu^{-}$
(Fig.~\ref{alldiff}a),
in $\Upsilon\left( 3\,{}^{3\!}S_{1}\right)$ $\to$
$\pi^{+}\pi^{-}\Upsilon\left( 2\,{}^{3\!}S_{1}\right)$
$\to$ $\pi^{+}\pi^{-}\mu^{+}\mu^{-}$
(Fig.~\ref{alldiff}b),
and in the $e^{+}e^{-}$ mass distributions of
$e^{+}e^{-}$ $\to$ $\pi^{+}\pi^{-}e^{+}e^{-}$
(Fig.~\ref{alldiff}c),
and finally the resonance signal shown in Fig.~\ref{hybrid},
is much too small to make firm claims.
However, we observe here that all points in the same direction.
Indeed, the probability must be close to zero that one accidentally finds
the same oscillations in four different sets of data
(Refs.~\cite{PRD79p111501R,ARXIV10095191}) involving different flavors,
statistical fluctuations at $\pm 38$ MeV
in yet another four sets of different data
(Figs.~\ref{mumuAll2S} and \ref{alldiff}),
and moreover a resonance-like fluctuation at $38$ MeV
in a further set of data (Fig.~\ref{hybrid}).
Furthermore, the related mass comes where predicted
by our analyses in mesonic spectroscopy
(see Ref.~\cite{ARXIV10112360} and references therein).

In Sec.~\ref{gluons}
we discussed that, most probably, the missing signal
is due to the emission
of an --- as yet unobserved --- light scalar particle,
while part of the excess data corroborates
such an interpretation.
Since the corresponding particle has all the right properties,
we conclude that we found first indications,
of the possible existence of a Higgs-like
particle, namely the scalar boson related to confinement.
Furthermore, the data also suggest the existence
of two replicas of the $E(38)$
with masses that are two and three times heavier than the $E(38)$.

In addition, we believe that this 38 MeV boson, which we
designate by $E(38)$, consists of a mini-universe filled with glue,
thus forming a very light scalar glueball.
Furthermore, we have pinpointed the masses
of possible $b\bar{b}g$ hybrids,
one of which shows up as an enhancement
in the invariant-mass distribution of BABAR data,
albeit with a 2$\sigma$ significance at most.

Finally, we urge the BABAR Collaboration
to inspect their larger data set
in order to settle, with higher statistics,
the possible existence of the $E(38)$
and the related $b\bar{b}g$ hybrid spectrum.

\section*{Acknowledgments}

We are grateful for the precise measurements
and data analyses of the BABAR, CDF, and CMD-2 Collaborations,
which made the present analysis possible.
One of us (EvB) wishes to thank
Drs. B.~Hiller, A.~H.~Blin, and A.~A.~Osipov
for useful discussions.
This work was supported in part by the {\it Funda\c{c}\~{a}o para a
Ci\^{e}ncia e a Tecnologia} \/of the {\it Minist\'{e}rio da Ci\^{e}ncia,
Tecnologia e Ensino Superior} \/of Portugal, under contract
CERN/\-FP/ 109307/\-2009.

\newcommand{\pubprt}[4]{#1 {\bf #2}, #3 (#4)}
\newcommand{\ertbid}[4]{[Erratum-ibid.~#1 {\bf #2}, #3 (#4)]}
\def\AdP{Annalen der Physik}
\def\AP{Ann.\ Phys.}
\def\EPJC{Eur.\ Phys.\ J.\ C}
\def\JETPL{JETP Lett.}
\def\MPLA{Mod.\ Phys.\ Lett.\ A}
\def\NCA{Nuovo Cim.\ A}
\def\PLB{Phys.\ Lett.\ B}
\def\PRD{Phys.\ Rev.\ D}
\def\PRL{Phys.\ Rev.\ Lett.}
\def\PREP{Phys.\ Rept.}
\def\SPAWB{Sitzungsber.\ Preuss.\ Akad.\ Wiss.\ Berlin (Math.\ Phys.\ )}
\def\ZPC{Z.\ Phys.\ C}

\end{document}